\pdfoutput=1

\documentclass[twocolumn,prl,reprint,preprintnumbers,amsmath,amssymb,superscriptaddress,nofootinbib]{revtex4}

\usepackage{aas_macros}

\usepackage{epsfig,psfrag,graphics,verbatim}
\usepackage{graphicx}
\usepackage{dcolumn}
\usepackage{bm}
\usepackage{color}

\begin{document}
\preprint{TUM-HEP 989/15}

\title{Testing modified Newtonian dynamics in the Milky Way}

\author{Fabio Iocco}
\affiliation{ICTP South American Institute for Fundamental Research, and Instituto de F\'isica Te\'orica - Universidade Estadual Paulista (UNESP), Rua Dr. Bento Teobaldo Ferraz 271, 01140-070 S\~{a}o Paulo, SP Brazil}
\affiliation{Instituto de F\'isica Te\'orica UAM/CSIC, C/ Nicol\'as Cabrera 13-15, 28049 Cantoblanco, Madrid, Spain}

\author{Miguel Pato}
\affiliation{The Oskar Klein Centre for Cosmoparticle Physics, Department of Physics, Stockholm University, AlbaNova, SE-106 91 Stockholm, Sweden}
\affiliation{Physik-Department T30d, Technische Universit\"at M\"unchen, James-Franck-Stra\ss{}e, D-85748 Garching, Germany}

\author{Gianfranco Bertone}
\affiliation{GRAPPA Institute, University of Amsterdam, Science Park 904, 1090 GL Amsterdam, The Netherlands}

\date{\today}

\begin{abstract}
Modified Newtonian dynamics (MOND) is an empirical theory originally proposed to explain the rotation curves of spiral galaxies by modifying the gravitational acceleration, rather than by invoking dark matter. Here,we set constraints on MOND using an up-to-date compilation of kinematic tracers of the Milky Way and a comprehensive collection of morphologies of the baryonic component in the Galaxy. In particular, we find that the so-called ``standard'' interpolating function cannot explain at the same time the rotation curve of the Milky Way and that of external galaxies for any of the baryonic models studied, while the so-called ``simple'' interpolating function can for a subset of models. Upcoming astronomical observations will refine our knowledge on the morphology of baryons and will ultimately confirm or rule out the validity of MOND in the Milky Way. We also present constraints on MOND-like theories without making any assumptions on the interpolating function.
\end{abstract}

\maketitle


\par If Newtonian gravity holds, the rotation curve of the Milky Way cannot be explained by visible (baryonic) matter only, therefore providing evidence for dark matter in the Galaxy, from its outskirts (e.g.~\cite{2003A&A...397..899S,2008ApJ...684.1143X,2014ApJ...794...59K}) down to inside the solar circle \cite{2015NatPh..11..245I}. In this paper we derive constraints on modifications of Newtonian gravity based on the same data set as used in Ref.~\cite{2015NatPh..11..245I}. We focus our attention on modified Newtonian dynamics (MOND), originally proposed by Milgrom \cite{1983ApJ...270..365M,1983ApJ...270..371M,1983ApJ...270..384M} (see also \cite{2002ARA&A..40..263S,2004PhRvD..70h3509B,2006ConPh..47..387B,2012LRR....15...10F,2014MNRAS.437.2531M}). MOND  postulates that below a characteristic value $a_0$ the acceleration $a$ is modified with respect to that predicted by Newton's law $a_N$ by the introduction of an interpolating function $\mu$ such that \cite{1983ApJ...270..365M,1983ApJ...270..371M}
\begin{equation}\label{mondeq}
\mu \left( \frac{a}{a_0} \right) a = a_N \, ,
\end{equation}
where $\mu(x) \simeq x$ for $x \ll 1$ and $\mu(x)\simeq 1$ for $x \gg 1$ (for modern theories of MOND, see \cite{2012LRR....15...10F}). The theory does not predict the value of $a_0$ nor a specific functional form for $\mu$ from first principles, so different proposals have been made in the literature (see e.g.~\cite{2006ConPh..47..387B,2012LRR....15...10F}). Two widely discussed functional forms for $\mu$ are the ``standard'' interpolating function,
\begin{equation}\label{mustd}
\mu_{\text{std}}(x) = \frac{x}{\sqrt{1+x^2}} \, ,
\end{equation}
and the ``simple'' interpolating function,
\begin{equation}\label{musim}
\mu_{\text{sim}}(x) = \frac{x}{1+x} \, .
\end{equation}

\par The limit of large accelerations, i.e.~$a \gg a_0$ or $x \gg 1$, is strongly constrained by solar system tests, where MOND is bound to recover Newtonian gravity. The limit of small accelerations, i.e.~$a \ll a_0$ or $x \ll 1$, is usually invoked in MOND studies to highlight how the observed flat rotation curves of spiral galaxies can be reproduced in the absence of dark matter.

\par In this paper, we set out to test the most common MOND scenarios with the latest data on the rotation curve of the Galaxy. We use a comprehensive compilation of kinematic tracers of the Milky Way and a state-of-the-art modelling of the baryons, both presented in Ref.~\cite{2015NatPh..11..245I}. Our results and analysis technique are complementary to previous MOND analyses of Milky Way data \cite{FamaeyBinney,2008ApJ...683..137M,2014ApJ...794..151L,MoffatToth} (see also \cite{2007MNRAS.379..597N,2009A&A...500..801B,Bovy:2013raa} for constraints from the vertical force in the discs of galaxies, including our own). The quantitative study of alternative formulations of MOND or other theories of modified gravity is left for future work.

\begin{figure*}[t]
\centering
\includegraphics[width=1.0\textwidth]{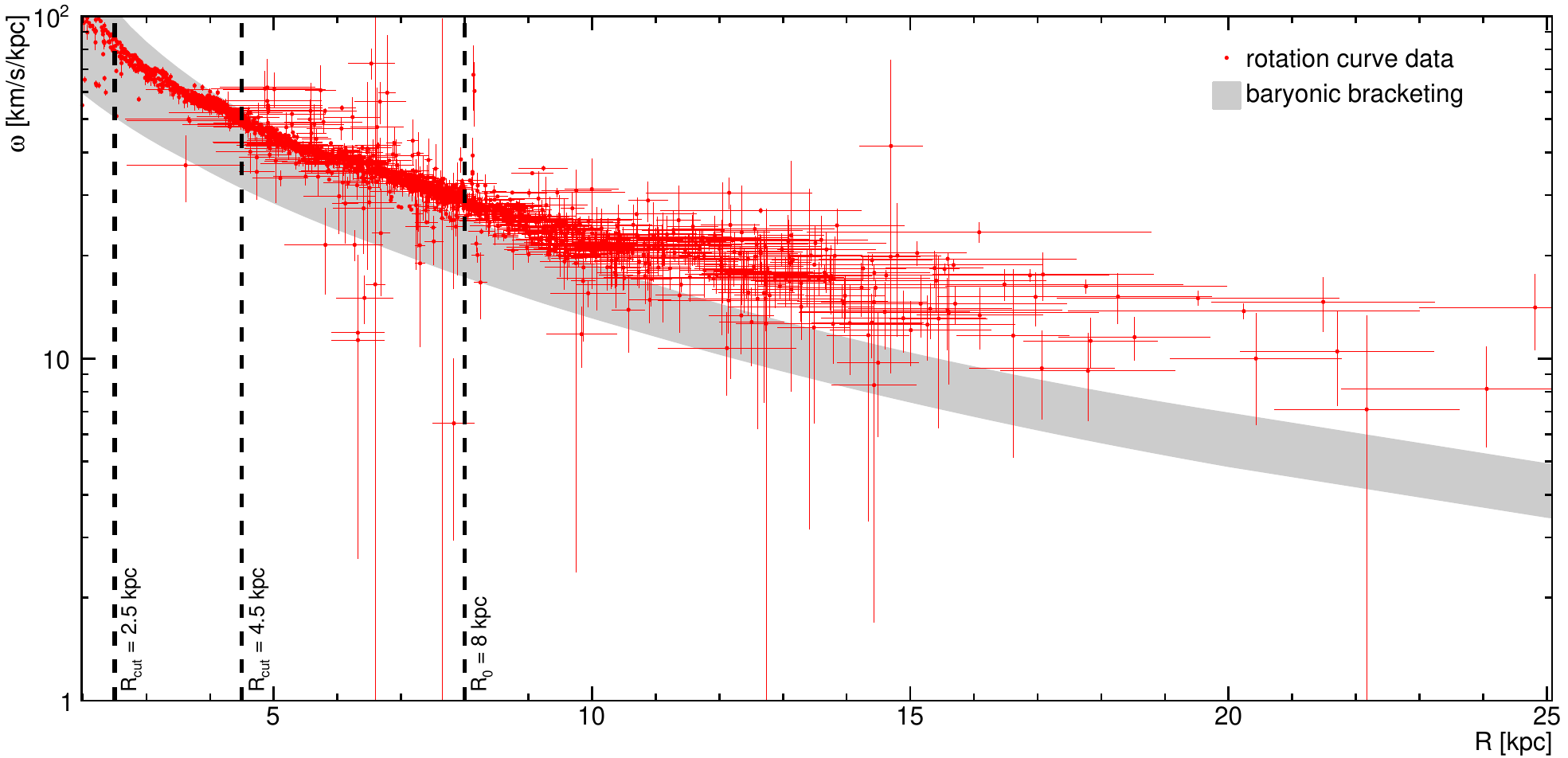}
\caption{The rotation curve of our Galaxy. The red data points indicate the angular circular velocity of all tracers in the rotation curve compilation, while the grey band brackets the contribution of all baryonic models under the assumption of Newtonian gravity. For further details, see Ref.~\cite{2015NatPh..11..245I}. We have adopted here $R_0=8\,$kpc, $v_0=230\,$km/s and $\left(U,V,W\right)_{\odot}=(11.10,12.24,7.25)\,$km/s \cite{Schoenrich2010}.}\label{fig:rotcurve}  
\end{figure*}

\par There are two observational inputs required to test MOND in our Galaxy: the observed gravitational acceleration $a$ and that predicted by Newtonian gravity $a_N$. These are the two key ingredients of our analysis. The acceleration $a$ is obtained from the rotation curve through $a=R\omega_c^2$, where $R$ is the Galactocentric radius and the angular circular velocity $\omega_c$ is set by the compilation of rotation curve measurements of Ref.~\cite{2015NatPh..11..245I}. This is a comprehensive compilation of kinematic tracers optimised to $R=3-20\,$kpc and that includes an unprecedented set of published data on gas kinematics, star kinematics and masers. A detailed description of the data used and their treatment can be found in the supplementary information of Ref.~\cite{2015NatPh..11..245I} (cf.~Tab.~S1 therein). The Newtonian acceleration $a_N=R\omega_{\textrm{b}}^2$ is set by the three-dimensional density distribution of baryons in our Galaxy for which we adopt the survey of observation-based models presented in Ref.~\cite{2015NatPh..11..245I}. This survey includes seven alternative morphologies for the stellar bulge \cite{Stanek1997,Zhao1996,BissantzGerhard2002,Vanhollebeke2009,LopezCorredoira2007,Robin2012}, five for the stellar disc \cite{HanGould2003,CalchiNovatiMancini2011,deJong2010,Juric2008,Bovy:2013raa} and two for the gas \cite{Ferriere1998,Moskalenko2002}, while the normalisation is set by microlensing observations for the bulge \cite{MACHO2005,2011JCAP...11..029I}, the local total stellar surface density for the disc \cite{Bovy:2013raa} and the CO-to-H$_2$ factor for the gas \cite{Ferriere2007,Ackermann2012}. For further details, please refer to the supplementary information of Ref.~\cite{2015NatPh..11..245I}. We use the morphologies of bulge, disc and gas in all 70 possible combinations, thus including all configurations of the baryonic component of the Galaxy present in the literature. In this way, we bracket the current uncertainty due to baryonic modelling and our conclusions do not rely on a specific model of the visible Galaxy but are solid against baryonic systematics. We adopt a distance to the Galactic centre $R_0=8\,$kpc, a local circular velocity $v_0=230\,$km/s and the peculiar solar motion $\left(U,V,W\right)_{\odot}=(11.10,12.24,7.25)\,$km/s \cite{Schoenrich2010}. The impact of varying these Galactic parameters is quantified later on, showing that our conclusions are robust against current uncertainties. Only kinematic tracers with Galactocentric radii above $R_{\textrm{cut}}=2.5\,$kpc are considered, amounting to $N=2686,\,2687,\,2715$ individual measurements for $R_0=7.98,\,8,\,8.68\,$kpc, respectively (cf.~below for the adopted uncertainty on $R_0$). The effect of pushing the radius cut to $R_{\textrm{cut}}=4.5\,$kpc is also quantified in order to avoid any influence of the bar (see e.g.~Ref.~\cite{2015A&A...578A..14C}); in this case, $N=2159,\,2162,\,2267$ for $R_0=7.98,\,8,\,8.68\,$kpc, respectively.

\par Fig.~\ref{fig:rotcurve} shows $\omega_c$, i.e.~the observed rotation curve, and $\omega_\textrm{b}$, i.e.~the rotation curve expected from baryons under Newtonian gravity. We are now in place to compare these two observation-based quantities and set constraints on MOND by assessing the discrepancy from the Newtonian scenario. The rotation curve predicted by a given MOND theory with fixed $\mu(x)$ and $a_0$ is obtained by solving Eq.~\eqref{mondeq} for $a=R\omega^2_{\textrm{mond}\color{black}}$. We then quantify the goodness-of-fit of the obtained $\omega_{\textrm{mond}\color{black}}(R)$ curve using a two-dimensional chi-square against the rotation curve data $\omega_c(R)$. The use of this test statistic is mainly motivated by the sizeable uncertainties on both $\omega_c$ and $R$ (see Ref.~\cite{2015NatPh..11..245I} for a technical discussion).

\begin{figure*}[t]
\centering
\includegraphics[width=0.49\textwidth]{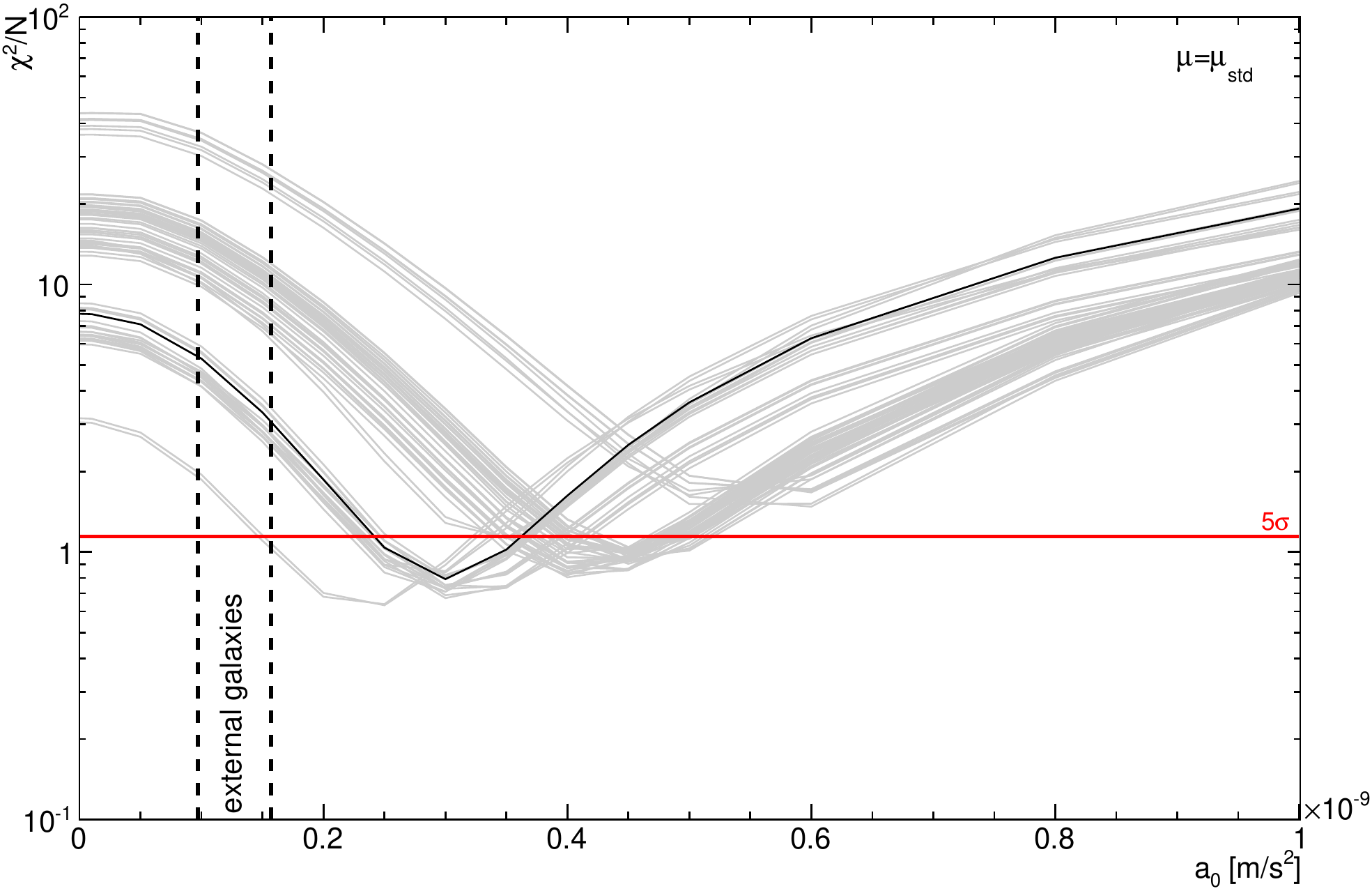}
\includegraphics[width=0.49\textwidth]{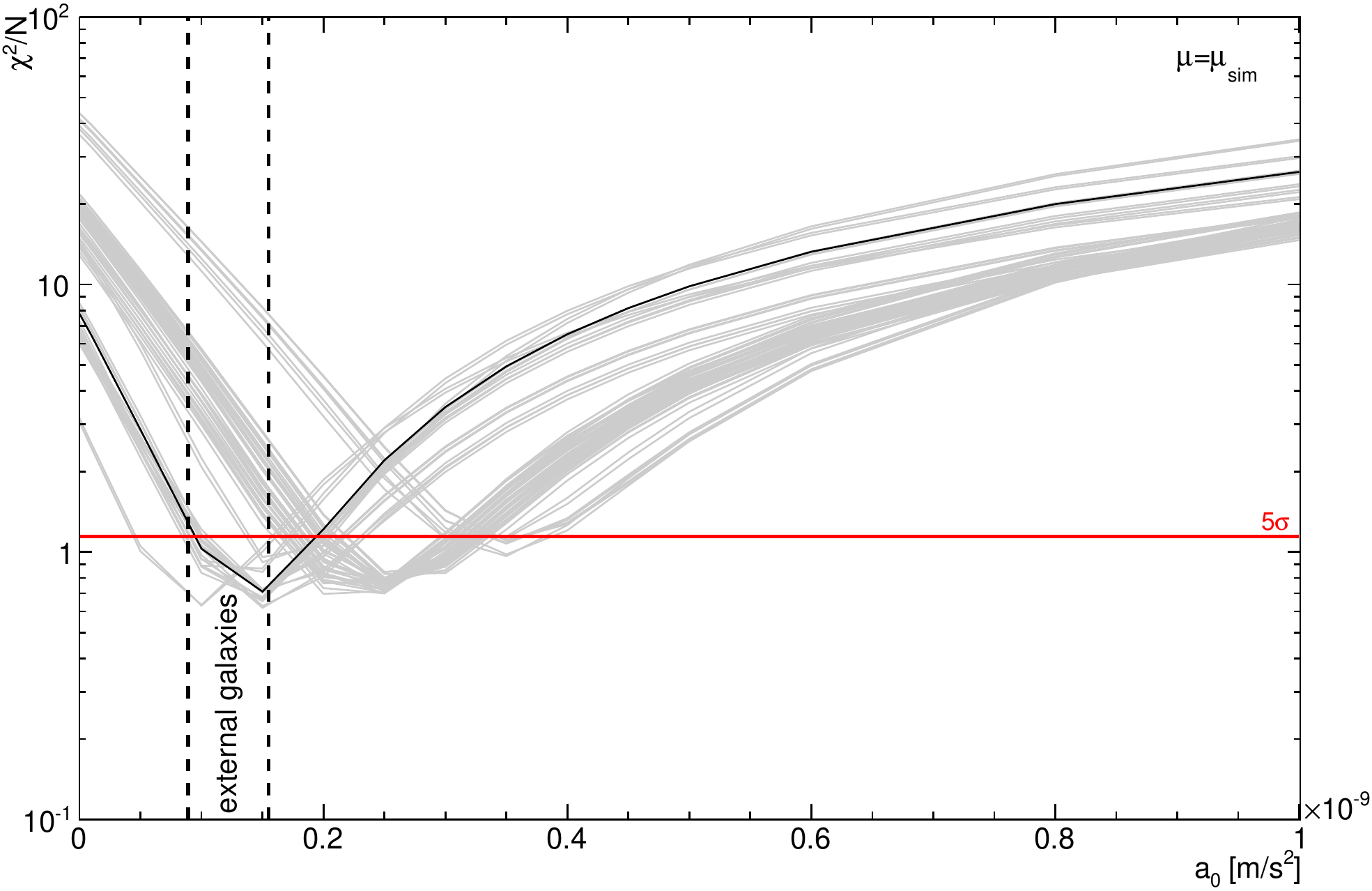}
\includegraphics[width=0.49\textwidth]{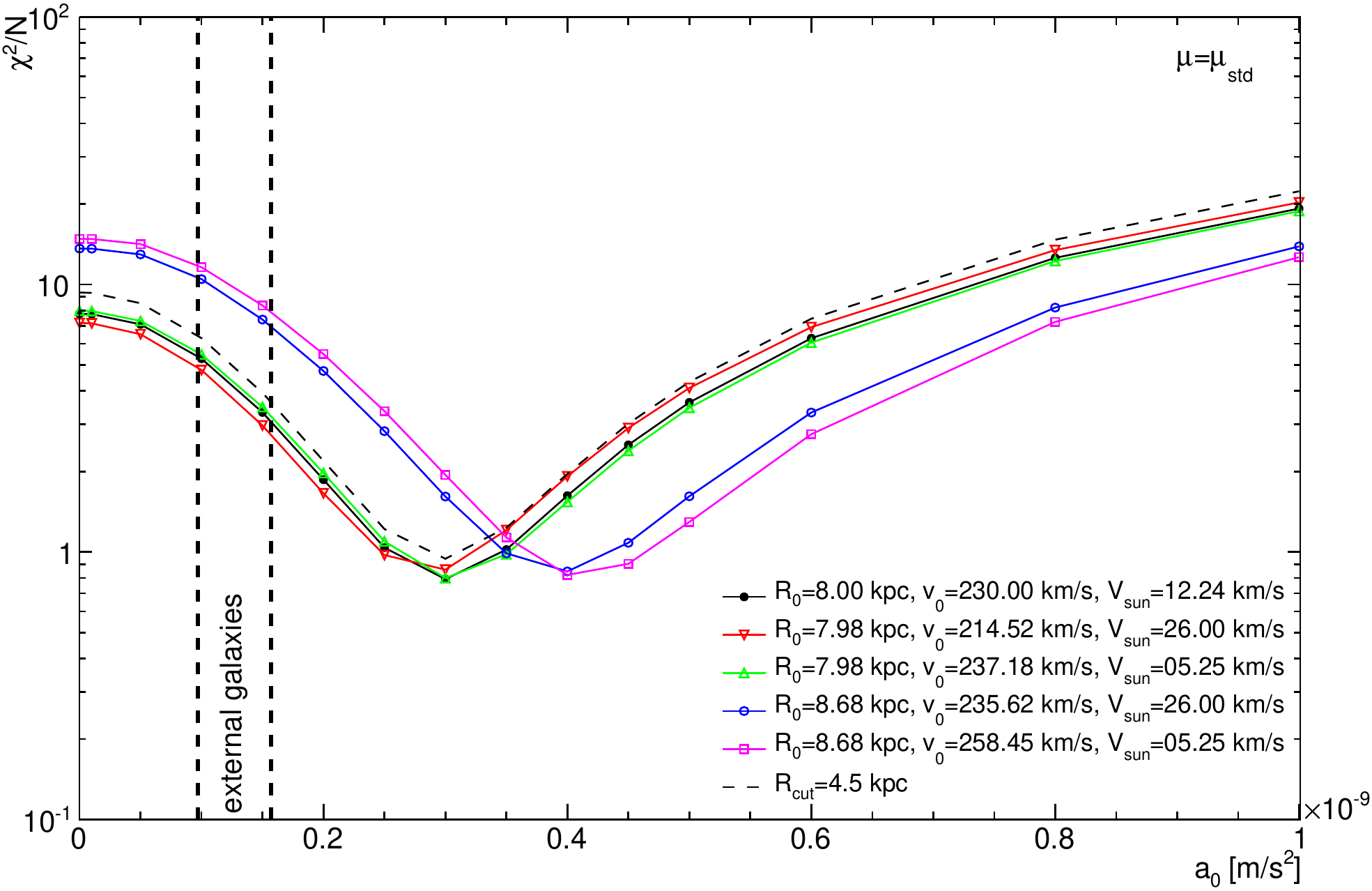}
\includegraphics[width=0.49\textwidth]{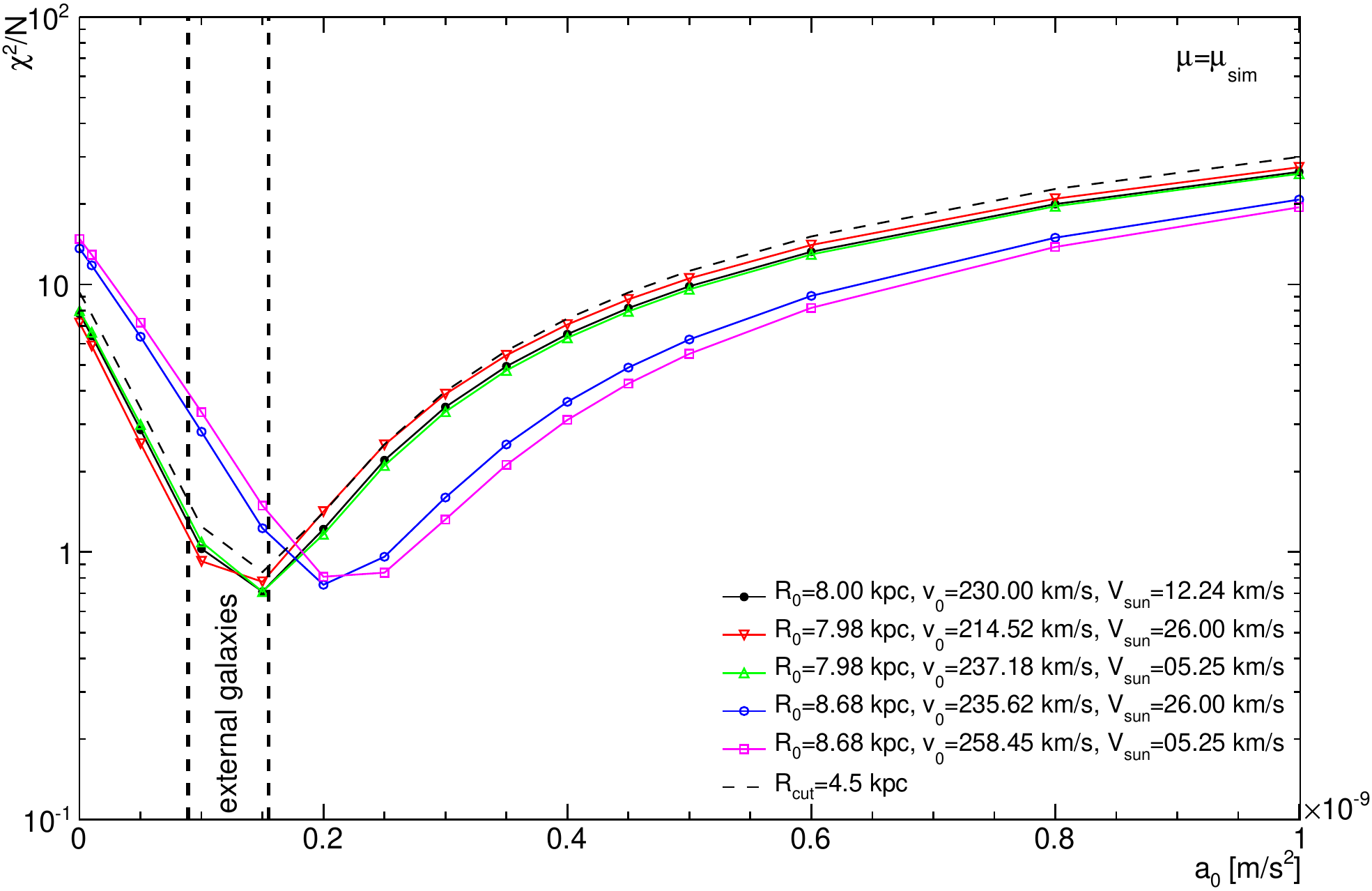}
\caption{Fitting MOND to the rotation curve of our Galaxy. The left (right) panels show the reduced chi-square of the MOND scenario with the standard (simple) interpolating function for different values of $a_0$. The upper panels convey the results for all baryonic models adopting $R_0=8\,$kpc, $v_0=230\,$km/s and $\left(U,V,W\right)_{\odot}=(11.10,12.24,7.25)\,$km/s \cite{Schoenrich2010}. The black line in each upper panel marks the reduced chi-square obtained for fiducial baryonic model I \cite{Stanek1997,Bovy:2013raa,Ferriere1998}. The bottom panels correspond to fiducial baryonic model I \cite{Stanek1997,Bovy:2013raa,Ferriere1998} with different combinations of Galactic parameters and radius cut. The range of $a_0$ found for each interpolating function from the rotation curves of external galaxies \cite{2011A&A...527A..76G} is encompassed by the vertical dashed lines, while the thick red line indicates the reduced chi-square corresponding to a $5\sigma$ exclusion.}
\label{fig:MONDfitting}  
\end{figure*}

\par The reduced chi-square as a function of the assumed $a_0$ is shown in the upper panels of Fig.~\ref{fig:MONDfitting} for each baryonic model for $\mu=\mu_{\text{std}}$ (left) and $\mu=\mu_{\text{sim}}$ (right). The results for what we shall call fiducial baryonic model I \cite{Stanek1997,Bovy:2013raa,Ferriere1998} are highlighted in black. Also shown are the values of $a_0$ favoured by the observation of rotation curves in external galaxies, namely $a_0=(1.27\pm 0.30) \times 10^{-10}\,$m/s$^2$ for $\mu_{\text{std}}$ and $a_0=(1.22\pm 0.33) \times 10^{-10}\,$m/s$^2$ for $\mu_{\text{sim}}$ \cite{2011A&A...527A..76G}, as well as the 5$\sigma$ exclusion line to guide the eye. In the case of the standard interpolating function, an acceptable best fit is obtained for most baryonic configurations. However, even in those cases, the preferred values of $a_0$ lie in the range $(2.5-5.0)\times 10^{-10}\,$m/s$^2$, clearly above the typical values found in external galaxies, which disfavours the standard interpolating function as being able to accommodate the rotation curve of external galaxies and that of the Milky Way at the same time. Instead, in the case of the simple interpolating function, the best-fit values of $a_0$ drop significantly to $(1.0-3.5)\times 10^{-10}\,$m/s$^2$, in line with the range inferred from external galaxies for a subset of baryonic models. Nevertheless, most baryonic models still prefer somewhat large values of $a_0$. These results are roughly in line with previous fits of MOND to Milky Way data \cite{FamaeyBinney,2008ApJ...683..137M}, but here we have expanded upon such analyses by using a comprehensive compilation of the available kinematic data and a wide range of baryonic models.

\begin{figure*}[t]
\centering
\includegraphics[width=1.0\textwidth]{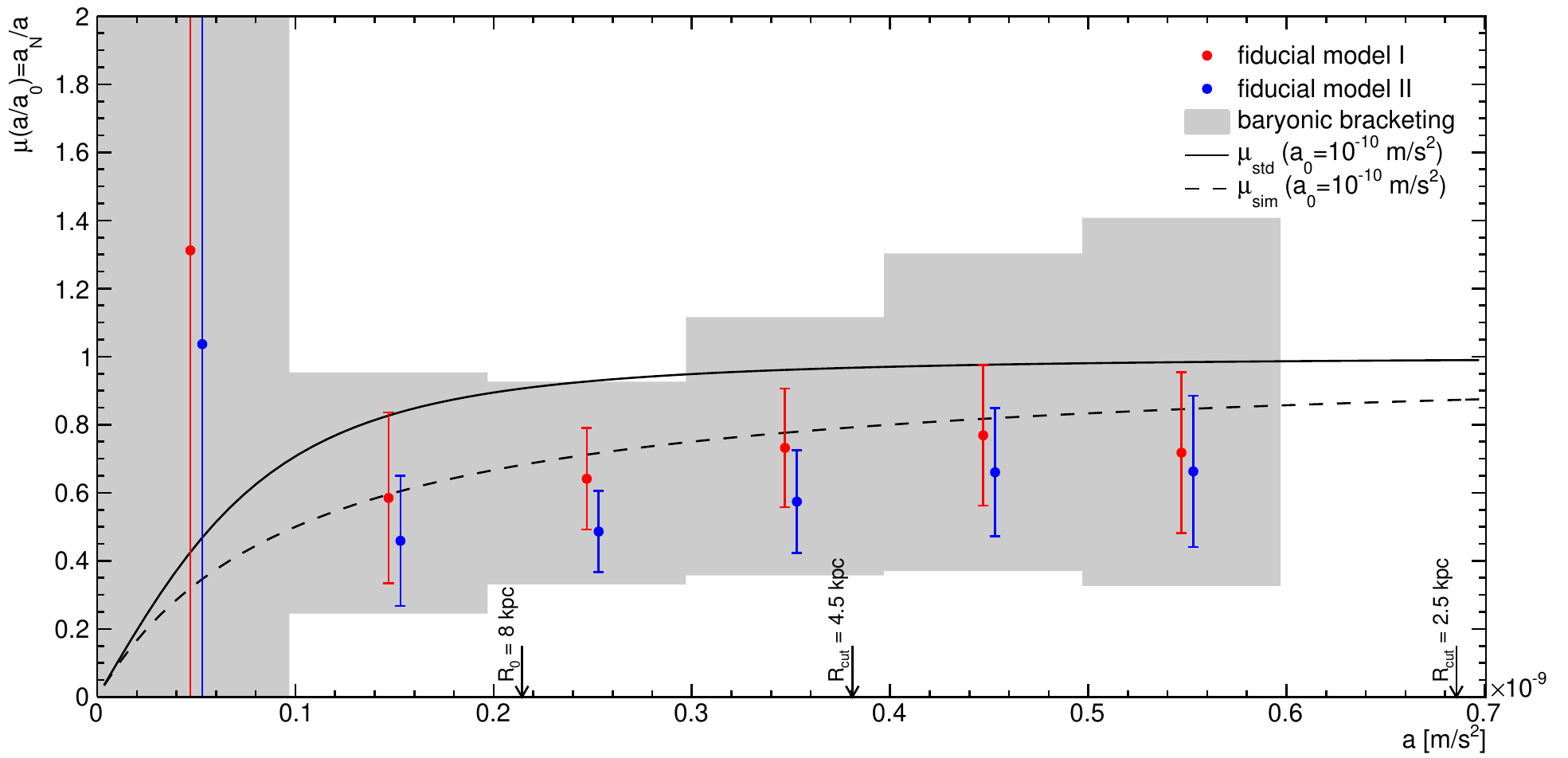}
\caption{The MOND interpolating function as inferred directly from the rotation curve of our Galaxy. The red and blue bars represent the binned 1$\sigma$ measurement of $\mu(a/a_0)=a_N/a$ for fiducial baryonic models I \cite{Stanek1997,Bovy:2013raa,Ferriere1998} and II \cite{Stanek1997,HanGould2003,Ferriere1998}, respectively, while the grey band encompasses the 1$\sigma$ measurements of all baryonic models. The slight shift in $a$ of the red and blue bars is for visualisation purposes only. Overplotted are the standard and simple interpolating functions for $a_0=10^{-10}\,\textrm{m/s}^2$. For reference, we also indicate the values of $a$ corresponding to $R_0$, $R_{\textrm{cut}}=2.5\,$kpc and $R_{\textrm{cut}}=4.5\,$kpc in the case of a flat rotation curve $v_c=v_0$. We have adopted here $R_0=8\,$kpc, $v_0=230\,$km/s and $\left(U,V,W\right)_{\odot}=(11.10,12.24,7.25)\,$km/s \cite{Schoenrich2010}.}
\label{fig:MONDmu}  
\end{figure*}

\par So far we have kept fixed $R_0=8\,$kpc, $v_0=230\,$km/s and $V_{\odot}=12.24\,$km/s. These values lie well within the ranges encompassing most current measurements, namely $R_0=8.0\pm0.5\,$kpc \cite{Gillessen2009,Ando2011,Malkin2012,Reid2014}, $v_0=230\pm20\,$km/s \cite{ReidBrunthaler2004,Reid2009,Bovy2009, McMillan2010, Bovy2012,Reid2014} and $V_{\odot}=5.25-26\,$km/s \cite{DehnenBinney1998,Schoenrich2010,Bovy2012,Reid2014}. It is important to understand how the uncertainties on $R_0$, $v_0$ and $V_{\odot}$ affect the results. For concreteness, let us focus on two specific measurements: (i) $R_0=8.33\pm0.35\,$kpc \cite{Gillessen2009}, provided by the monitoring of the orbits of S-stars around the central supermassive black hole, and (ii) $\Omega_{\odot}\equiv\frac{v_0+V_{\odot}}{R_0}=30.26\pm0.12\,\color{black}$km/s/kpc \cite{ReidBrunthaler2004}, provided by the proper motion of Sgr A$^{\ast}$. Combining (i) and (ii) with the allowed values of $V_\odot$ reported above, we have the following 1$\sigma$ configurations:\\
(a) $R_0=7.98\,$kpc, $v_0=214.52\,\color{black}$km/s for $V_{\odot}=26\,$km/s;\\ 
(b) $R_0=7.98\,$kpc, $v_0=237.18\,\color{black}$km/s for $V_{\odot}=5.25\,$km/s;\\ 
(c) $R_0=8.68\,$kpc, $v_0=235.62\,\color{black}$km/s for $V_{\odot}=26\,$km/s; \\ 
(d) $R_0=8.68\,$kpc, $v_0=258.45\,\color{black}$km/s for $V_{\odot}=5.25\,$km/s.\\ 
We use configurations (a) through (d) to quantify the impact of current uncertainties on our results. This is shown explicitly in the bottom panels of Fig.~\ref{fig:MONDfitting} for fiducial baryonic model I \cite{Stanek1997,Bovy:2013raa,Ferriere1998}. It is possible to conclude that, within the same baryonic model, the best-fit values of $a_0$ are relatively insensitive to the actual local circular velocity $v_0$ provided the Sgr A$^\ast$ constraint is met (cf.~(a) and (b)). Moreover, a value of $R_0$ on the high end of the currently allowed range shifts the favoured $a_0$ to higher accelerations, i.e.~away from the range inferred from external galaxies (cf.~(c) and (d)). These considerations hold for both $\mu_{\textrm{std}}$ and $\mu_{\textrm{sim}}$. Therefore, the conclusions drawn in the previous paragraphs do not change qualitatively when varying the Galatic fundamental parameters. The same applies when pushing the radius cut to $4.5\,$kpc, as also shown in the bottom panels of Fig.~\ref{fig:MONDfitting}.

\par In Fig.~\ref{fig:MONDmu} we perform a data-driven test for MOND-like theories that can be applied directly to rotation curve and photometric data. By mapping the modified acceleration with the rotation curve data through $a=R\omega_c^2$ and the Newtonian acceleration with the baryonic distribution through $a_N=R\omega_{\textrm{b}}^2$, we reconstruct $\mu$ directly from Eq.~\eqref{mondeq} with no dependence on $a_0$, namely $\mu(a/a_0)=a_N/a$. Then, the resulting measurements of $\mu$ are binned in linear intervals of size $\Delta a=10^{-10}\,\textrm{m/s}^2$ requiring at least five rotation curve measurements per bin. This is shown in Fig.~\ref{fig:MONDmu} by the red bars for fiducial baryonic model I \cite{Stanek1997,Bovy:2013raa,Ferriere1998}, by the blue bars for an additional baryonic model II \cite{Stanek1997,HanGould2003,Ferriere1998} and by the grey band for the bracketing of all baryonic models implemented in our analysis. In the same plot we superimpose $\mu_{\text{std}}(a/a_0)$ and $\mu_{\text{sim}}(a/a_0)$ for $a_0=10^{-10}\,\textrm{m/s}^2$. As clear from the figure, the standard interpolating function is in tension with the data for both fiducial baryonic models, whereas the simple interpolating function remains viable for fiducial baryonic model I. Note that for a flat rotation curve $v_c=v_0$ the acceleration reads $a=v_0^2/R$, so the kinematic tracers at small $R$ correspond to large $a$ and vice-versa. Therefore, the importance of baryonic modelling grows for large $a$ since that corresponds to the region of the Galaxy where the baryons become more and more important. Although at present uncertainties are still sizeable, Fig.~\ref{fig:MONDmu} provides a powerful non-parametric test of the MOND paradigm. Our analysis is complementary to that of Ref.~\cite{2014ApJ...794..151L}, which covers a slightly different acceleration range, and that of Ref.~\cite{MoffatToth}, which is more focussed on the outer Galaxy.

\par From Figs.~\ref{fig:MONDfitting} and \ref{fig:MONDmu}, it appears that the simple interpolating function provides a better fit to Milky Way data, in agreement with earlier studies which adopt a different setup and approach, e.g.~\cite{FamaeyBinney,2008ApJ...683..137M}. Let us recall that this functional form is strongly disfavoured by solar system tests and it has to be modified in the large acceleration limit (see e.g.~\cite{2011A&A...527A..76G,2012LRR....15...10F}). Modifications of the simple interpolating function have been proposed in the literature (i.e., the family of $\nu$-functions \cite{2012LRR....15...10F} or the ``improved simple'' function \cite{2011A&A...527A..76G}) by making $\mu$ converge to unity for $a/a_0\gtrsim 10$. We note that, for the range of accelerations to which we are sensitive (cf.~Fig.~\ref{fig:MONDmu}), these modified functions and the simple function are virtually indistinguishable for typical values of $a_0$ and the results presented here for $\mu_{\textrm{sim}}$ would also apply to those modified functions. As our measurements of the Galactic rotation curve and our understanding of the distribution of baryons improve, we expect these tests to provide more and more stringent constraints on both $\mu$ and $a_0$. For the time being, however, we can conservatively state -- based on an up-to-date compilation of rotation curve data and a comprehensive collection of data-inferred morphologies of the baryonic component -- that MOND variants employing the standard interpolating function do not fit simultaneously the rotation curves of external galaxies and that of the Milky Way.

\par Let us finally point out that we have adopted in our analysis a wide compilation of baryonic models to bracket the current uncertainty on the morphology and composition of our Galaxy. In the coming years, astronomical surveys such as Gaia \cite{2012Ap&SS.341...31D}, APOGEE-2 (SDSS-IV) \cite{apogee2site}, WFIRST \cite{2015arXiv150303757S}, WEAVE \cite{weavesite} and 4MOST \cite{2012SPIE.8446E..0TD} will eventually help narrow down on the baryonic distribution, thus providing a decisive step towards testing the MOND paradigm in the Milky Way.

\vspace{0.5cm}
{\it Acknowledgements.} The authors thank Beno\^{i}t Famaey for useful comments on the manuscript. F.~I.~acknowledges support from the Simons Foundation and FAPESP process 2014/22985-1, M.~P.~from Wenner-Gren Stiftelserna in Stockholm, and G.~B.~from the European Research Council through the ERC Starting Grant {\it WIMPs Kairos}.

\bibliographystyle{apsrev}
\bibliography{testmond}

\begin{thebibliography}{53}
\expandafter\ifx\csname natexlab\endcsname\relax\def\natexlab#1{#1}\fi
\expandafter\ifx\csname bibnamefont\endcsname\relax
  \def\bibnamefont#1{#1}\fi
\expandafter\ifx\csname bibfnamefont\endcsname\relax
  \def\bibfnamefont#1{#1}\fi
\expandafter\ifx\csname citenamefont\endcsname\relax
  \def\citenamefont#1{#1}\fi
\expandafter\ifx\csname url\endcsname\relax
  \def\url#1{\texttt{#1}}\fi
\expandafter\ifx\csname urlprefix\endcsname\relax\def\urlprefix{URL }\fi
\providecommand{\bibinfo}[2]{#2}
\providecommand{\eprint}[2][]{\url{#2}}

\bibitem[{\citenamefont{{Sakamoto} et~al.}(2003)\citenamefont{{Sakamoto},
  {Chiba}, and {Beers}}}]{2003A&A...397..899S}
\bibinfo{author}{\bibfnamefont{T.}~\bibnamefont{{Sakamoto}}},
  \bibinfo{author}{\bibfnamefont{M.}~\bibnamefont{{Chiba}}}, \bibnamefont{and}
  \bibinfo{author}{\bibfnamefont{T.~C.} \bibnamefont{{Beers}}},
  \bibinfo{journal}{\aap} \textbf{\bibinfo{volume}{397}}, \bibinfo{pages}{899}
  (\bibinfo{year}{2003}), \eprint{astro-ph/0210508}.

\bibitem[{\citenamefont{{Xue} et~al.}(2008)\citenamefont{{Xue}, {Rix}, {Zhao},
  {Re Fiorentin}, {Naab}, {Steinmetz}, {van den Bosch}, {Beers}, {Lee}, {Bell}
  et~al.}}]{2008ApJ...684.1143X}
\bibinfo{author}{\bibfnamefont{X.~X.} \bibnamefont{{Xue}}},
  \bibinfo{author}{\bibfnamefont{H.~W.} \bibnamefont{{Rix}}},
  \bibinfo{author}{\bibfnamefont{G.}~\bibnamefont{{Zhao}}},
  \bibinfo{author}{\bibfnamefont{P.}~\bibnamefont{{Re Fiorentin}}},
  \bibinfo{author}{\bibfnamefont{T.}~\bibnamefont{{Naab}}},
  \bibinfo{author}{\bibfnamefont{M.}~\bibnamefont{{Steinmetz}}},
  \bibinfo{author}{\bibfnamefont{F.~C.} \bibnamefont{{van den Bosch}}},
  \bibinfo{author}{\bibfnamefont{T.~C.} \bibnamefont{{Beers}}},
  \bibinfo{author}{\bibfnamefont{Y.~S.} \bibnamefont{{Lee}}},
  \bibinfo{author}{\bibfnamefont{E.~F.} \bibnamefont{{Bell}}},
  \bibnamefont{et~al.}, \bibinfo{journal}{\apj} \textbf{\bibinfo{volume}{684}},
  \bibinfo{pages}{1143} (\bibinfo{year}{2008}), \eprint{0801.1232}.

\bibitem[{\citenamefont{{Kafle} et~al.}(2014)\citenamefont{{Kafle}, {Sharma},
  {Lewis}, and {Bland-Hawthorn}}}]{2014ApJ...794...59K}
\bibinfo{author}{\bibfnamefont{P.~R.} \bibnamefont{{Kafle}}},
  \bibinfo{author}{\bibfnamefont{S.}~\bibnamefont{{Sharma}}},
  \bibinfo{author}{\bibfnamefont{G.~F.} \bibnamefont{{Lewis}}},
  \bibnamefont{and}
  \bibinfo{author}{\bibfnamefont{J.}~\bibnamefont{{Bland-Hawthorn}}},
  \bibinfo{journal}{\apj} \textbf{\bibinfo{volume}{794}}, \bibinfo{eid}{59}
  (\bibinfo{year}{2014}), \eprint{1408.1787}.

\bibitem[{\citenamefont{{Iocco} et~al.}(2015)\citenamefont{{Iocco}, {Pato}, and
  {Bertone}}}]{2015NatPh..11..245I}
\bibinfo{author}{\bibfnamefont{F.}~\bibnamefont{{Iocco}}},
  \bibinfo{author}{\bibfnamefont{M.}~\bibnamefont{{Pato}}}, \bibnamefont{and}
  \bibinfo{author}{\bibfnamefont{G.}~\bibnamefont{{Bertone}}},
  \bibinfo{journal}{Nature Physics} \textbf{\bibinfo{volume}{11}},
  \bibinfo{pages}{245} (\bibinfo{year}{2015}), \eprint{1502.03821}.

\bibitem[{\citenamefont{{Milgrom}}(1983{\natexlab{a}})}]{1983ApJ...270..365M}
\bibinfo{author}{\bibfnamefont{M.}~\bibnamefont{{Milgrom}}},
  \bibinfo{journal}{\apj} \textbf{\bibinfo{volume}{270}}, \bibinfo{pages}{365}
  (\bibinfo{year}{1983}{\natexlab{a}}).

\bibitem[{\citenamefont{{Milgrom}}(1983{\natexlab{b}})}]{1983ApJ...270..371M}
\bibinfo{author}{\bibfnamefont{M.}~\bibnamefont{{Milgrom}}},
  \bibinfo{journal}{\apj} \textbf{\bibinfo{volume}{270}}, \bibinfo{pages}{371}
  (\bibinfo{year}{1983}{\natexlab{b}}).

\bibitem[{\citenamefont{{Milgrom}}(1983{\natexlab{c}})}]{1983ApJ...270..384M}
\bibinfo{author}{\bibfnamefont{M.}~\bibnamefont{{Milgrom}}},
  \bibinfo{journal}{\apj} \textbf{\bibinfo{volume}{270}}, \bibinfo{pages}{384}
  (\bibinfo{year}{1983}{\natexlab{c}}).

\bibitem[{\citenamefont{{Sanders} and {McGaugh}}(2002)}]{2002ARA&A..40..263S}
\bibinfo{author}{\bibfnamefont{R.~H.} \bibnamefont{{Sanders}}}
  \bibnamefont{and} \bibinfo{author}{\bibfnamefont{S.~S.}
  \bibnamefont{{McGaugh}}}, \bibinfo{journal}{\araa}
  \textbf{\bibinfo{volume}{40}}, \bibinfo{pages}{263} (\bibinfo{year}{2002}),
  \eprint{astro-ph/0204521}.

\bibitem[{\citenamefont{{Bekenstein}}(2004)}]{2004PhRvD..70h3509B}
\bibinfo{author}{\bibfnamefont{J.~D.} \bibnamefont{{Bekenstein}}},
  \bibinfo{journal}{\prd} \textbf{\bibinfo{volume}{70}}, \bibinfo{eid}{083509}
  (\bibinfo{year}{2004}), \eprint{astro-ph/0403694}.

\bibitem[{\citenamefont{{Bekenstein}}(2006)}]{2006ConPh..47..387B}
\bibinfo{author}{\bibfnamefont{J.}~\bibnamefont{{Bekenstein}}},
  \bibinfo{journal}{Contemporary Physics} \textbf{\bibinfo{volume}{47}},
  \bibinfo{pages}{387} (\bibinfo{year}{2006}), \eprint{astro-ph/0701848}.

\bibitem[{\citenamefont{{Famaey} and {McGaugh}}(2012)}]{2012LRR....15...10F}
\bibinfo{author}{\bibfnamefont{B.}~\bibnamefont{{Famaey}}} \bibnamefont{and}
  \bibinfo{author}{\bibfnamefont{S.~S.} \bibnamefont{{McGaugh}}},
  \bibinfo{journal}{Living Reviews in Relativity}
  \textbf{\bibinfo{volume}{15}}, \bibinfo{pages}{10} (\bibinfo{year}{2012}),
  \eprint{1112.3960}.

\bibitem[{\citenamefont{{Milgrom}}(2014)}]{2014MNRAS.437.2531M}
\bibinfo{author}{\bibfnamefont{M.}~\bibnamefont{{Milgrom}}},
  \bibinfo{journal}{\mnras} \textbf{\bibinfo{volume}{437}},
  \bibinfo{pages}{2531} (\bibinfo{year}{2014}), \eprint{1212.2568}.

\bibitem[{\citenamefont{{Famaey} and {Binney}}(2005)}]{FamaeyBinney}
\bibinfo{author}{\bibfnamefont{B.}~\bibnamefont{{Famaey}}} \bibnamefont{and}
  \bibinfo{author}{\bibfnamefont{J.}~\bibnamefont{{Binney}}},
  \bibinfo{journal}{\mnras} \textbf{\bibinfo{volume}{363}},
  \bibinfo{pages}{603} (\bibinfo{year}{2005}), \eprint{astro-ph/0506723}.

\bibitem[{\citenamefont{{McGaugh}}(2008)}]{2008ApJ...683..137M}
\bibinfo{author}{\bibfnamefont{S.~S.} \bibnamefont{{McGaugh}}},
  \bibinfo{journal}{\apj} \textbf{\bibinfo{volume}{683}}, \bibinfo{pages}{137}
  (\bibinfo{year}{2008}), \eprint{0804.1314}.

\bibitem[{\citenamefont{{Loebman} et~al.}(2014)\citenamefont{{Loebman},
  {Ivezi{\'c}}, {Quinn}, {Bovy}, {Christensen}, {Juri{\'c}}, {Ro{\v s}kar},
  {Brooks}, and {Governato}}}]{2014ApJ...794..151L}
\bibinfo{author}{\bibfnamefont{S.~R.} \bibnamefont{{Loebman}}},
  \bibinfo{author}{\bibfnamefont{{\v Z}.}~\bibnamefont{{Ivezi{\'c}}}},
  \bibinfo{author}{\bibfnamefont{T.~R.} \bibnamefont{{Quinn}}},
  \bibinfo{author}{\bibfnamefont{J.}~\bibnamefont{{Bovy}}},
  \bibinfo{author}{\bibfnamefont{C.~R.} \bibnamefont{{Christensen}}},
  \bibinfo{author}{\bibfnamefont{M.}~\bibnamefont{{Juri{\'c}}}},
  \bibinfo{author}{\bibfnamefont{R.}~\bibnamefont{{Ro{\v s}kar}}},
  \bibinfo{author}{\bibfnamefont{A.~M.} \bibnamefont{{Brooks}}},
  \bibnamefont{and}
  \bibinfo{author}{\bibfnamefont{F.}~\bibnamefont{{Governato}}},
  \bibinfo{journal}{\apj} \textbf{\bibinfo{volume}{794}}, \bibinfo{eid}{151}
  (\bibinfo{year}{2014}), \eprint{1408.5388}.

\bibitem[{\citenamefont{{Moffat} and {Toth}}(2014)}]{MoffatToth}
\bibinfo{author}{\bibfnamefont{J.~W.} \bibnamefont{{Moffat}}} \bibnamefont{and}
  \bibinfo{author}{\bibfnamefont{V.~T.} \bibnamefont{{Toth}}},
  \bibinfo{journal}{ArXiv e-prints}  (\bibinfo{year}{2014}),
  \eprint{1411.6701}.

\bibitem[{\citenamefont{{Nipoti} et~al.}(2007)\citenamefont{{Nipoti},
  {Londrillo}, {Zhao}, and {Ciotti}}}]{2007MNRAS.379..597N}
\bibinfo{author}{\bibfnamefont{C.}~\bibnamefont{{Nipoti}}},
  \bibinfo{author}{\bibfnamefont{P.}~\bibnamefont{{Londrillo}}},
  \bibinfo{author}{\bibfnamefont{H.}~\bibnamefont{{Zhao}}}, \bibnamefont{and}
  \bibinfo{author}{\bibfnamefont{L.}~\bibnamefont{{Ciotti}}},
  \bibinfo{journal}{\mnras} \textbf{\bibinfo{volume}{379}},
  \bibinfo{pages}{597} (\bibinfo{year}{2007}), \eprint{0704.0740}.

\bibitem[{\citenamefont{{Bienaym{\'e}}
  et~al.}(2009)\citenamefont{{Bienaym{\'e}}, {Famaey}, {Wu}, {Zhao}, and
  {Aubert}}}]{2009A&A...500..801B}
\bibinfo{author}{\bibfnamefont{O.}~\bibnamefont{{Bienaym{\'e}}}},
  \bibinfo{author}{\bibfnamefont{B.}~\bibnamefont{{Famaey}}},
  \bibinfo{author}{\bibfnamefont{X.}~\bibnamefont{{Wu}}},
  \bibinfo{author}{\bibfnamefont{H.~S.} \bibnamefont{{Zhao}}},
  \bibnamefont{and} \bibinfo{author}{\bibfnamefont{D.}~\bibnamefont{{Aubert}}},
  \bibinfo{journal}{\aap} \textbf{\bibinfo{volume}{500}}, \bibinfo{pages}{801}
  (\bibinfo{year}{2009}), \eprint{0904.3893}.

\bibitem[{\citenamefont{{Bovy} and {Rix}}(2013)}]{Bovy:2013raa}
\bibinfo{author}{\bibfnamefont{J.}~\bibnamefont{{Bovy}}} \bibnamefont{and}
  \bibinfo{author}{\bibfnamefont{H.-W.} \bibnamefont{{Rix}}},
  \bibinfo{journal}{\apj} \textbf{\bibinfo{volume}{779}}, \bibinfo{eid}{115}
  (\bibinfo{year}{2013}), \eprint{1309.0809}.

\bibitem[{\citenamefont{{Sch{\"o}nrich}
  et~al.}(2010)\citenamefont{{Sch{\"o}nrich}, {Binney}, and
  {Dehnen}}}]{Schoenrich2010}
\bibinfo{author}{\bibfnamefont{R.}~\bibnamefont{{Sch{\"o}nrich}}},
  \bibinfo{author}{\bibfnamefont{J.}~\bibnamefont{{Binney}}}, \bibnamefont{and}
  \bibinfo{author}{\bibfnamefont{W.}~\bibnamefont{{Dehnen}}},
  \bibinfo{journal}{\mnras} \textbf{\bibinfo{volume}{403}},
  \bibinfo{pages}{1829} (\bibinfo{year}{2010}), \eprint{0912.3693}.

\bibitem[{\citenamefont{{Stanek} et~al.}(1997)\citenamefont{{Stanek},
  {Udalski}, {Szymanski}, {Kaluzny}, {Kubiak}, {Mateo}, and
  {Krzeminski}}}]{Stanek1997}
\bibinfo{author}{\bibfnamefont{K.~Z.} \bibnamefont{{Stanek}}},
  \bibinfo{author}{\bibfnamefont{A.}~\bibnamefont{{Udalski}}},
  \bibinfo{author}{\bibfnamefont{M.}~\bibnamefont{{Szymanski}}},
  \bibinfo{author}{\bibfnamefont{J.}~\bibnamefont{{Kaluzny}}},
  \bibinfo{author}{\bibfnamefont{M.}~\bibnamefont{{Kubiak}}},
  \bibinfo{author}{\bibfnamefont{M.}~\bibnamefont{{Mateo}}}, \bibnamefont{and}
  \bibinfo{author}{\bibfnamefont{W.}~\bibnamefont{{Krzeminski}}},
  \bibinfo{journal}{\apj} \textbf{\bibinfo{volume}{477}}, \bibinfo{pages}{163}
  (\bibinfo{year}{1997}), \eprint{astro-ph/9605162}.

\bibitem[{\citenamefont{{Zhao}}(1996)}]{Zhao1996}
\bibinfo{author}{\bibfnamefont{H.}~\bibnamefont{{Zhao}}},
  \bibinfo{journal}{\mnras} \textbf{\bibinfo{volume}{283}},
  \bibinfo{pages}{149} (\bibinfo{year}{1996}), \eprint{astro-ph/9512064}.

\bibitem[{\citenamefont{{Bissantz} and {Gerhard}}(2002)}]{BissantzGerhard2002}
\bibinfo{author}{\bibfnamefont{N.}~\bibnamefont{{Bissantz}}} \bibnamefont{and}
  \bibinfo{author}{\bibfnamefont{O.}~\bibnamefont{{Gerhard}}},
  \bibinfo{journal}{\mnras} \textbf{\bibinfo{volume}{330}},
  \bibinfo{pages}{591} (\bibinfo{year}{2002}), \eprint{astro-ph/0110368}.

\bibitem[{\citenamefont{{Vanhollebeke}
  et~al.}(2009)\citenamefont{{Vanhollebeke}, {Groenewegen}, and
  {Girardi}}}]{Vanhollebeke2009}
\bibinfo{author}{\bibfnamefont{E.}~\bibnamefont{{Vanhollebeke}}},
  \bibinfo{author}{\bibfnamefont{M.~A.~T.} \bibnamefont{{Groenewegen}}},
  \bibnamefont{and}
  \bibinfo{author}{\bibfnamefont{L.}~\bibnamefont{{Girardi}}},
  \bibinfo{journal}{\aap} \textbf{\bibinfo{volume}{498}}, \bibinfo{pages}{95}
  (\bibinfo{year}{2009}).

\bibitem[{\citenamefont{{L{\'o}pez-Corredoira}
  et~al.}(2007)\citenamefont{{L{\'o}pez-Corredoira}, {Cabrera-Lavers},
  {Mahoney}, {Hammersley}, {Garz{\'o}n}, and
  {Gonz{\'a}lez-Fern{\'a}ndez}}}]{LopezCorredoira2007}
\bibinfo{author}{\bibfnamefont{M.}~\bibnamefont{{L{\'o}pez-Corredoira}}},
  \bibinfo{author}{\bibfnamefont{A.}~\bibnamefont{{Cabrera-Lavers}}},
  \bibinfo{author}{\bibfnamefont{T.~J.} \bibnamefont{{Mahoney}}},
  \bibinfo{author}{\bibfnamefont{P.~L.} \bibnamefont{{Hammersley}}},
  \bibinfo{author}{\bibfnamefont{F.}~\bibnamefont{{Garz{\'o}n}}},
  \bibnamefont{and}
  \bibinfo{author}{\bibfnamefont{C.}~\bibnamefont{{Gonz{\'a}lez-Fern{\'a}ndez}%
}}, \bibinfo{journal}{\aj} \textbf{\bibinfo{volume}{133}}, \bibinfo{pages}{154}
  (\bibinfo{year}{2007}), \eprint{astro-ph/0606201}.

\bibitem[{\citenamefont{{Robin} et~al.}(2012)\citenamefont{{Robin}, {Marshall},
  {Schultheis}, and {Reyl{\'e}}}}]{Robin2012}
\bibinfo{author}{\bibfnamefont{A.~C.} \bibnamefont{{Robin}}},
  \bibinfo{author}{\bibfnamefont{D.~J.} \bibnamefont{{Marshall}}},
  \bibinfo{author}{\bibfnamefont{M.}~\bibnamefont{{Schultheis}}},
  \bibnamefont{and}
  \bibinfo{author}{\bibfnamefont{C.}~\bibnamefont{{Reyl{\'e}}}},
  \bibinfo{journal}{\aap} \textbf{\bibinfo{volume}{538}}, \bibinfo{eid}{A106}
  (\bibinfo{year}{2012}), \eprint{1111.5744}.

\bibitem[{\citenamefont{{Han} and {Gould}}(2003)}]{HanGould2003}
\bibinfo{author}{\bibfnamefont{C.}~\bibnamefont{{Han}}} \bibnamefont{and}
  \bibinfo{author}{\bibfnamefont{A.}~\bibnamefont{{Gould}}},
  \bibinfo{journal}{\apj} \textbf{\bibinfo{volume}{592}}, \bibinfo{pages}{172}
  (\bibinfo{year}{2003}), \eprint{astro-ph/0303309}.

\bibitem[{\citenamefont{{Calchi Novati} and
  {Mancini}}(2011)}]{CalchiNovatiMancini2011}
\bibinfo{author}{\bibfnamefont{S.}~\bibnamefont{{Calchi Novati}}}
  \bibnamefont{and}
  \bibinfo{author}{\bibfnamefont{L.}~\bibnamefont{{Mancini}}},
  \bibinfo{journal}{\mnras} \textbf{\bibinfo{volume}{416}},
  \bibinfo{pages}{1292} (\bibinfo{year}{2011}), \eprint{1105.4615}.

\bibitem[{\citenamefont{{de Jong} et~al.}(2010)\citenamefont{{de Jong},
  {Yanny}, {Rix}, {Dolphin}, {Martin}, and {Beers}}}]{deJong2010}
\bibinfo{author}{\bibfnamefont{J.~T.~A.} \bibnamefont{{de Jong}}},
  \bibinfo{author}{\bibfnamefont{B.}~\bibnamefont{{Yanny}}},
  \bibinfo{author}{\bibfnamefont{H.-W.} \bibnamefont{{Rix}}},
  \bibinfo{author}{\bibfnamefont{A.~E.} \bibnamefont{{Dolphin}}},
  \bibinfo{author}{\bibfnamefont{N.~F.} \bibnamefont{{Martin}}},
  \bibnamefont{and} \bibinfo{author}{\bibfnamefont{T.~C.}
  \bibnamefont{{Beers}}}, \bibinfo{journal}{\apj}
  \textbf{\bibinfo{volume}{714}}, \bibinfo{pages}{663} (\bibinfo{year}{2010}),
  \eprint{0911.3900}.

\bibitem[{\citenamefont{{Juri{\'c}} et~al.}(2008)\citenamefont{{Juri{\'c}},
  {Ivezi{\'c}}, {Brooks}, {Lupton}, {Schlegel}, {Finkbeiner}, {Padmanabhan},
  {Bond}, {Sesar}, {Rockosi} et~al.}}]{Juric2008}
\bibinfo{author}{\bibfnamefont{M.}~\bibnamefont{{Juri{\'c}}}},
  \bibinfo{author}{\bibfnamefont{{\v Z}.}~\bibnamefont{{Ivezi{\'c}}}},
  \bibinfo{author}{\bibfnamefont{A.}~\bibnamefont{{Brooks}}},
  \bibinfo{author}{\bibfnamefont{R.~H.} \bibnamefont{{Lupton}}},
  \bibinfo{author}{\bibfnamefont{D.}~\bibnamefont{{Schlegel}}},
  \bibinfo{author}{\bibfnamefont{D.}~\bibnamefont{{Finkbeiner}}},
  \bibinfo{author}{\bibfnamefont{N.}~\bibnamefont{{Padmanabhan}}},
  \bibinfo{author}{\bibfnamefont{N.}~\bibnamefont{{Bond}}},
  \bibinfo{author}{\bibfnamefont{B.}~\bibnamefont{{Sesar}}},
  \bibinfo{author}{\bibfnamefont{C.~M.} \bibnamefont{{Rockosi}}},
  \bibnamefont{et~al.}, \bibinfo{journal}{\apj} \textbf{\bibinfo{volume}{673}},
  \bibinfo{pages}{864} (\bibinfo{year}{2008}), \eprint{astro-ph/0510520}.

\bibitem[{\citenamefont{{Ferriere}}(1998)}]{Ferriere1998}
\bibinfo{author}{\bibfnamefont{K.}~\bibnamefont{{Ferriere}}},
  \bibinfo{journal}{\apj} \textbf{\bibinfo{volume}{497}}, \bibinfo{pages}{759}
  (\bibinfo{year}{1998}).

\bibitem[{\citenamefont{{Moskalenko} et~al.}(2002)\citenamefont{{Moskalenko},
  {Strong}, {Ormes}, and {Potgieter}}}]{Moskalenko2002}
\bibinfo{author}{\bibfnamefont{I.~V.} \bibnamefont{{Moskalenko}}},
  \bibinfo{author}{\bibfnamefont{A.~W.} \bibnamefont{{Strong}}},
  \bibinfo{author}{\bibfnamefont{J.~F.} \bibnamefont{{Ormes}}},
  \bibnamefont{and} \bibinfo{author}{\bibfnamefont{M.~S.}
  \bibnamefont{{Potgieter}}}, \bibinfo{journal}{\apj}
  \textbf{\bibinfo{volume}{565}}, \bibinfo{pages}{280} (\bibinfo{year}{2002}),
  \eprint{astro-ph/0106567}.

\bibitem[{\citenamefont{{Popowski} et~al.}(2005)\citenamefont{{Popowski},
  {Griest}, {Thomas}, {Cook}, {Bennett}, {Becker}, {Alves}, {Minniti}, {Drake},
  {Alcock} et~al.}}]{MACHO2005}
\bibinfo{author}{\bibfnamefont{P.}~\bibnamefont{{Popowski}}},
  \bibinfo{author}{\bibfnamefont{K.}~\bibnamefont{{Griest}}},
  \bibinfo{author}{\bibfnamefont{C.~L.} \bibnamefont{{Thomas}}},
  \bibinfo{author}{\bibfnamefont{K.~H.} \bibnamefont{{Cook}}},
  \bibinfo{author}{\bibfnamefont{D.~P.} \bibnamefont{{Bennett}}},
  \bibinfo{author}{\bibfnamefont{A.~C.} \bibnamefont{{Becker}}},
  \bibinfo{author}{\bibfnamefont{D.~R.} \bibnamefont{{Alves}}},
  \bibinfo{author}{\bibfnamefont{D.}~\bibnamefont{{Minniti}}},
  \bibinfo{author}{\bibfnamefont{A.~J.} \bibnamefont{{Drake}}},
  \bibinfo{author}{\bibfnamefont{C.}~\bibnamefont{{Alcock}}},
  \bibnamefont{et~al.}, \bibinfo{journal}{\apj} \textbf{\bibinfo{volume}{631}},
  \bibinfo{pages}{879} (\bibinfo{year}{2005}), \eprint{astro-ph/0410319}.

\bibitem[{\citenamefont{{Iocco} et~al.}(2011)\citenamefont{{Iocco}, {Pato},
  {Bertone}, and {Jetzer}}}]{2011JCAP...11..029I}
\bibinfo{author}{\bibfnamefont{F.}~\bibnamefont{{Iocco}}},
  \bibinfo{author}{\bibfnamefont{M.}~\bibnamefont{{Pato}}},
  \bibinfo{author}{\bibfnamefont{G.}~\bibnamefont{{Bertone}}},
  \bibnamefont{and} \bibinfo{author}{\bibfnamefont{P.}~\bibnamefont{{Jetzer}}},
  \bibinfo{journal}{\jcap} \textbf{\bibinfo{volume}{11}}, \bibinfo{eid}{029}
  (\bibinfo{year}{2011}), \eprint{1107.5810}.

\bibitem[{\citenamefont{{Ferri{\`e}re}
  et~al.}(2007)\citenamefont{{Ferri{\`e}re}, {Gillard}, and
  {Jean}}}]{Ferriere2007}
\bibinfo{author}{\bibfnamefont{K.}~\bibnamefont{{Ferri{\`e}re}}},
  \bibinfo{author}{\bibfnamefont{W.}~\bibnamefont{{Gillard}}},
  \bibnamefont{and} \bibinfo{author}{\bibfnamefont{P.}~\bibnamefont{{Jean}}},
  \bibinfo{journal}{\aap} \textbf{\bibinfo{volume}{467}}, \bibinfo{pages}{611}
  (\bibinfo{year}{2007}).

\bibitem[{\citenamefont{{Ackermann} et~al.}(2012)\citenamefont{{Ackermann},
  {Ajello}, {Atwood}, {Baldini}, {Ballet}, {Barbiellini}, {Bastieri},
  {Bechtol}, {Bellazzini}, {Berenji} et~al.}}]{Ackermann2012}
\bibinfo{author}{\bibfnamefont{M.}~\bibnamefont{{Ackermann}}},
  \bibinfo{author}{\bibfnamefont{M.}~\bibnamefont{{Ajello}}},
  \bibinfo{author}{\bibfnamefont{W.~B.} \bibnamefont{{Atwood}}},
  \bibinfo{author}{\bibfnamefont{L.}~\bibnamefont{{Baldini}}},
  \bibinfo{author}{\bibfnamefont{J.}~\bibnamefont{{Ballet}}},
  \bibinfo{author}{\bibfnamefont{G.}~\bibnamefont{{Barbiellini}}},
  \bibinfo{author}{\bibfnamefont{D.}~\bibnamefont{{Bastieri}}},
  \bibinfo{author}{\bibfnamefont{K.}~\bibnamefont{{Bechtol}}},
  \bibinfo{author}{\bibfnamefont{R.}~\bibnamefont{{Bellazzini}}},
  \bibinfo{author}{\bibfnamefont{B.}~\bibnamefont{{Berenji}}},
  \bibnamefont{et~al.}, \bibinfo{journal}{\apj} \textbf{\bibinfo{volume}{750}},
  \bibinfo{eid}{3} (\bibinfo{year}{2012}).

\bibitem[{\citenamefont{{Chemin} et~al.}(2015)\citenamefont{{Chemin}, {Renaud},
  and {Soubiran}}}]{2015A&A...578A..14C}
\bibinfo{author}{\bibfnamefont{L.}~\bibnamefont{{Chemin}}},
  \bibinfo{author}{\bibfnamefont{F.}~\bibnamefont{{Renaud}}}, \bibnamefont{and}
  \bibinfo{author}{\bibfnamefont{C.}~\bibnamefont{{Soubiran}}},
  \bibinfo{journal}{\aap} \textbf{\bibinfo{volume}{578}}, \bibinfo{eid}{A14}
  (\bibinfo{year}{2015}), \eprint{1504.01507}.

\bibitem[{\citenamefont{{Gentile} et~al.}(2011)\citenamefont{{Gentile},
  {Famaey}, and {de Blok}}}]{2011A&A...527A..76G}
\bibinfo{author}{\bibfnamefont{G.}~\bibnamefont{{Gentile}}},
  \bibinfo{author}{\bibfnamefont{B.}~\bibnamefont{{Famaey}}}, \bibnamefont{and}
  \bibinfo{author}{\bibfnamefont{W.~J.~G.} \bibnamefont{{de Blok}}},
  \bibinfo{journal}{\aap} \textbf{\bibinfo{volume}{527}}, \bibinfo{eid}{A76}
  (\bibinfo{year}{2011}), \eprint{1011.4148}.

\bibitem[{\citenamefont{{Gillessen} et~al.}(2009)\citenamefont{{Gillessen},
  {Eisenhauer}, {Trippe}, {Alexander}, {Genzel}, {Martins}, and
  {Ott}}}]{Gillessen2009}
\bibinfo{author}{\bibfnamefont{S.}~\bibnamefont{{Gillessen}}},
  \bibinfo{author}{\bibfnamefont{F.}~\bibnamefont{{Eisenhauer}}},
  \bibinfo{author}{\bibfnamefont{S.}~\bibnamefont{{Trippe}}},
  \bibinfo{author}{\bibfnamefont{T.}~\bibnamefont{{Alexander}}},
  \bibinfo{author}{\bibfnamefont{R.}~\bibnamefont{{Genzel}}},
  \bibinfo{author}{\bibfnamefont{F.}~\bibnamefont{{Martins}}},
  \bibnamefont{and} \bibinfo{author}{\bibfnamefont{T.}~\bibnamefont{{Ott}}},
  \bibinfo{journal}{\apj} \textbf{\bibinfo{volume}{692}}, \bibinfo{pages}{1075}
  (\bibinfo{year}{2009}).

\bibitem[{\citenamefont{{Ando} et~al.}(2011)\citenamefont{{Ando}, {Nagayama},
  {Omodaka}, {Handa}, {Imai}, {Nakagawa}, {Nakanishi}, {Honma}, {Kobayashi},
  and {Miyaji}}}]{Ando2011}
\bibinfo{author}{\bibfnamefont{K.}~\bibnamefont{{Ando}}},
  \bibinfo{author}{\bibfnamefont{T.}~\bibnamefont{{Nagayama}}},
  \bibinfo{author}{\bibfnamefont{T.}~\bibnamefont{{Omodaka}}},
  \bibinfo{author}{\bibfnamefont{T.}~\bibnamefont{{Handa}}},
  \bibinfo{author}{\bibfnamefont{H.}~\bibnamefont{{Imai}}},
  \bibinfo{author}{\bibfnamefont{A.}~\bibnamefont{{Nakagawa}}},
  \bibinfo{author}{\bibfnamefont{H.}~\bibnamefont{{Nakanishi}}},
  \bibinfo{author}{\bibfnamefont{M.}~\bibnamefont{{Honma}}},
  \bibinfo{author}{\bibfnamefont{H.}~\bibnamefont{{Kobayashi}}},
  \bibnamefont{and} \bibinfo{author}{\bibfnamefont{T.}~\bibnamefont{{Miyaji}}},
  \bibinfo{journal}{\pasj} \textbf{\bibinfo{volume}{63}}, \bibinfo{pages}{45}
  (\bibinfo{year}{2011}).

\bibitem[{\citenamefont{{Malkin}}(2012)}]{Malkin2012}
\bibinfo{author}{\bibfnamefont{Z.}~\bibnamefont{{Malkin}}},
  \bibinfo{journal}{ArXiv e-prints}  (\bibinfo{year}{2012}),
  \eprint{1202.6128}.

\bibitem[{\citenamefont{{Reid} et~al.}(2014)\citenamefont{{Reid}, {Menten},
  {Brunthaler}, {Zheng}, {Dame}, {Xu}, {Wu}, {Zhang}, {Sanna}, {Sato}
  et~al.}}]{Reid2014}
\bibinfo{author}{\bibfnamefont{M.~J.} \bibnamefont{{Reid}}},
  \bibinfo{author}{\bibfnamefont{K.~M.} \bibnamefont{{Menten}}},
  \bibinfo{author}{\bibfnamefont{A.}~\bibnamefont{{Brunthaler}}},
  \bibinfo{author}{\bibfnamefont{X.~W.} \bibnamefont{{Zheng}}},
  \bibinfo{author}{\bibfnamefont{T.~M.} \bibnamefont{{Dame}}},
  \bibinfo{author}{\bibfnamefont{Y.}~\bibnamefont{{Xu}}},
  \bibinfo{author}{\bibfnamefont{Y.}~\bibnamefont{{Wu}}},
  \bibinfo{author}{\bibfnamefont{B.}~\bibnamefont{{Zhang}}},
  \bibinfo{author}{\bibfnamefont{A.}~\bibnamefont{{Sanna}}},
  \bibinfo{author}{\bibfnamefont{M.}~\bibnamefont{{Sato}}},
  \bibnamefont{et~al.}, \bibinfo{journal}{\apj} \textbf{\bibinfo{volume}{783}},
  \bibinfo{eid}{130} (\bibinfo{year}{2014}).

\bibitem[{\citenamefont{{Reid} and {Brunthaler}}(2004)}]{ReidBrunthaler2004}
\bibinfo{author}{\bibfnamefont{M.~J.} \bibnamefont{{Reid}}} \bibnamefont{and}
  \bibinfo{author}{\bibfnamefont{A.}~\bibnamefont{{Brunthaler}}},
  \bibinfo{journal}{\apj} \textbf{\bibinfo{volume}{616}}, \bibinfo{pages}{872}
  (\bibinfo{year}{2004}), \eprint{astro-ph/0408107}.

\bibitem[{\citenamefont{{Reid} et~al.}(2009)\citenamefont{{Reid}, {Menten},
  {Zheng}, {Brunthaler}, {Moscadelli}, {Xu}, {Zhang}, {Sato}, {Honma}, {Hirota}
  et~al.}}]{Reid2009}
\bibinfo{author}{\bibfnamefont{M.~J.} \bibnamefont{{Reid}}},
  \bibinfo{author}{\bibfnamefont{K.~M.} \bibnamefont{{Menten}}},
  \bibinfo{author}{\bibfnamefont{X.~W.} \bibnamefont{{Zheng}}},
  \bibinfo{author}{\bibfnamefont{A.}~\bibnamefont{{Brunthaler}}},
  \bibinfo{author}{\bibfnamefont{L.}~\bibnamefont{{Moscadelli}}},
  \bibinfo{author}{\bibfnamefont{Y.}~\bibnamefont{{Xu}}},
  \bibinfo{author}{\bibfnamefont{B.}~\bibnamefont{{Zhang}}},
  \bibinfo{author}{\bibfnamefont{M.}~\bibnamefont{{Sato}}},
  \bibinfo{author}{\bibfnamefont{M.}~\bibnamefont{{Honma}}},
  \bibinfo{author}{\bibfnamefont{T.}~\bibnamefont{{Hirota}}},
  \bibnamefont{et~al.}, \bibinfo{journal}{\apj} \textbf{\bibinfo{volume}{700}},
  \bibinfo{eid}{137} (\bibinfo{year}{2009}).

\bibitem[{\citenamefont{{Bovy} et~al.}(2009)\citenamefont{{Bovy}, {Hogg}, and
  {Rix}}}]{Bovy2009}
\bibinfo{author}{\bibfnamefont{J.}~\bibnamefont{{Bovy}}},
  \bibinfo{author}{\bibfnamefont{D.~W.} \bibnamefont{{Hogg}}},
  \bibnamefont{and} \bibinfo{author}{\bibfnamefont{H.-W.} \bibnamefont{{Rix}}},
  \bibinfo{journal}{\apj} \textbf{\bibinfo{volume}{704}}, \bibinfo{pages}{1704}
  (\bibinfo{year}{2009}).

\bibitem[{\citenamefont{{McMillan} and {Binney}}(2010)}]{McMillan2010}
\bibinfo{author}{\bibfnamefont{P.~J.} \bibnamefont{{McMillan}}}
  \bibnamefont{and} \bibinfo{author}{\bibfnamefont{J.~J.}
  \bibnamefont{{Binney}}}, \bibinfo{journal}{\mnras}
  \textbf{\bibinfo{volume}{402}}, \bibinfo{pages}{934} (\bibinfo{year}{2010}).

\bibitem[{\citenamefont{{Bovy} et~al.}(2012)\citenamefont{{Bovy}, {Allende
  Prieto}, {Beers}, {Bizyaev}, {da Costa}, {Cunha}, {Ebelke}, {Eisenstein},
  {Frinchaboy}, {Garc{\'{\i}}a P{\'e}rez} et~al.}}]{Bovy2012}
\bibinfo{author}{\bibfnamefont{J.}~\bibnamefont{{Bovy}}},
  \bibinfo{author}{\bibfnamefont{C.}~\bibnamefont{{Allende Prieto}}},
  \bibinfo{author}{\bibfnamefont{T.~C.} \bibnamefont{{Beers}}},
  \bibinfo{author}{\bibfnamefont{D.}~\bibnamefont{{Bizyaev}}},
  \bibinfo{author}{\bibfnamefont{L.~N.} \bibnamefont{{da Costa}}},
  \bibinfo{author}{\bibfnamefont{K.}~\bibnamefont{{Cunha}}},
  \bibinfo{author}{\bibfnamefont{G.~L.} \bibnamefont{{Ebelke}}},
  \bibinfo{author}{\bibfnamefont{D.~J.} \bibnamefont{{Eisenstein}}},
  \bibinfo{author}{\bibfnamefont{P.~M.} \bibnamefont{{Frinchaboy}}},
  \bibinfo{author}{\bibfnamefont{A.~E.} \bibnamefont{{Garc{\'{\i}}a
  P{\'e}rez}}}, \bibnamefont{et~al.}, \bibinfo{journal}{\apj}
  \textbf{\bibinfo{volume}{759}}, \bibinfo{eid}{131} (\bibinfo{year}{2012}).

\bibitem[{\citenamefont{{Dehnen} and {Binney}}(1998)}]{DehnenBinney1998}
\bibinfo{author}{\bibfnamefont{W.}~\bibnamefont{{Dehnen}}} \bibnamefont{and}
  \bibinfo{author}{\bibfnamefont{J.~J.} \bibnamefont{{Binney}}},
  \bibinfo{journal}{\mnras} \textbf{\bibinfo{volume}{298}},
  \bibinfo{pages}{387} (\bibinfo{year}{1998}).

\bibitem[{\citenamefont{{de Bruijne}}(2012)}]{2012Ap&SS.341...31D}
\bibinfo{author}{\bibfnamefont{J.~H.~J.} \bibnamefont{{de Bruijne}}},
  \bibinfo{journal}{\apss} \textbf{\bibinfo{volume}{341}}, \bibinfo{pages}{31}
  (\bibinfo{year}{2012}), \eprint{1201.3238}.

\bibitem[{apo()}]{apogee2site}
\bibinfo{howpublished}{\url{http://www.sdss.org/surveys/apogee-2/}}.

\bibitem[{\citenamefont{{Spergel} et~al.}(2015)\citenamefont{{Spergel},
  {Gehrels}, {Baltay}, {Bennett}, {Breckinridge}, {Donahue}, {Dressler},
  {Gaudi}, {Greene}, {Guyon} et~al.}}]{2015arXiv150303757S}
\bibinfo{author}{\bibfnamefont{D.}~\bibnamefont{{Spergel}}},
  \bibinfo{author}{\bibfnamefont{N.}~\bibnamefont{{Gehrels}}},
  \bibinfo{author}{\bibfnamefont{C.}~\bibnamefont{{Baltay}}},
  \bibinfo{author}{\bibfnamefont{D.}~\bibnamefont{{Bennett}}},
  \bibinfo{author}{\bibfnamefont{J.}~\bibnamefont{{Breckinridge}}},
  \bibinfo{author}{\bibfnamefont{M.}~\bibnamefont{{Donahue}}},
  \bibinfo{author}{\bibfnamefont{A.}~\bibnamefont{{Dressler}}},
  \bibinfo{author}{\bibfnamefont{B.~S.} \bibnamefont{{Gaudi}}},
  \bibinfo{author}{\bibfnamefont{T.}~\bibnamefont{{Greene}}},
  \bibinfo{author}{\bibfnamefont{O.}~\bibnamefont{{Guyon}}},
  \bibnamefont{et~al.}, \bibinfo{journal}{ArXiv e-prints}
  (\bibinfo{year}{2015}), \eprint{1503.03757}.

\bibitem[{wea()}]{weavesite}
\bibinfo{howpublished}{\url{http://www.ing.iac.es/weave/index.html}}.

\bibitem[{\citenamefont{{de Jong} et~al.}(2012)\citenamefont{{de Jong},
  {Bellido-Tirado}, {Chiappini}, {Depagne}, {Haynes}, {Johl}, {Schnurr},
  {Schwope}, {Walcher}, {Dionies} et~al.}}]{2012SPIE.8446E..0TD}
\bibinfo{author}{\bibfnamefont{R.~S.} \bibnamefont{{de Jong}}},
  \bibinfo{author}{\bibfnamefont{O.}~\bibnamefont{{Bellido-Tirado}}},
  \bibinfo{author}{\bibfnamefont{C.}~\bibnamefont{{Chiappini}}},
  \bibinfo{author}{\bibfnamefont{{\'E}.}~\bibnamefont{{Depagne}}},
  \bibinfo{author}{\bibfnamefont{R.}~\bibnamefont{{Haynes}}},
  \bibinfo{author}{\bibfnamefont{D.}~\bibnamefont{{Johl}}},
  \bibinfo{author}{\bibfnamefont{O.}~\bibnamefont{{Schnurr}}},
  \bibinfo{author}{\bibfnamefont{A.}~\bibnamefont{{Schwope}}},
  \bibinfo{author}{\bibfnamefont{J.}~\bibnamefont{{Walcher}}},
  \bibinfo{author}{\bibfnamefont{F.}~\bibnamefont{{Dionies}}},
  \bibnamefont{et~al.}, in \emph{\bibinfo{booktitle}{Society of Photo-Optical
  Instrumentation Engineers (SPIE) Conference Series}} (\bibinfo{year}{2012}),
  vol. \bibinfo{volume}{8446} of \emph{\bibinfo{series}{Society of
  Photo-Optical Instrumentation Engineers (SPIE) Conference Series}},
  p.~\bibinfo{pages}{0}, \eprint{1206.6885}.

\end{thebibliography}


\end{document}